# Security Risks of Porting C Programs to WebAssembly

Quentin Stiévenart
Vrije Universiteit Brussel
Belgium
quentin.stievenart@vub.be

Coen De Roover
Vrije Universiteit Brussel
Belgium
coen.de.roover@vub.be

Mohammad Ghafari
University of Auckland
New Zealand
m.ghafari@auckland.ac.nz

## ABSTRACT

WebAssembly is a compilation target for cross-platform applications that is increasingly being used. In this paper, we investigate whether one can transparently cross-compile C programs to WebAssembly, and if not, what impact porting can have on their security. We compile 17 802 programs that exhibit common vulnerabilities to 64-bit x86 and to WebAssembly binaries, and we observe that the execution of 4 911 binaries produces different results across these platforms. Through manual inspection, we identify three classes of root causes for such differences: the use of a different standard library implementation, the lack of security measures in WebAssembly, and the different semantics of the execution environments. We describe our observations and discuss the ones that are critical from a security point of view and need most attention from developers. We conclude that compiling an existing C program to WebAssembly for cross-platform distribution may require source code adaptations; otherwise, the security of the WebAssembly application may be at risk.

## 1 INTRODUCTION

WebAssembly is a recent standard for portable binary code that aims to bring native speed to programs that run in web browsers [12]. Major browsers support WebAssembly, and its adoption has expanded to, for example, IoT systems and cross-platform desktop applications [1, 13]. It is also supported as a compilation target by compilers such as Clang and Emscripten, enabling the compilation of a wide variety of source languages to WebAssembly.

Now that it suffices to compile to WebAssembly, it seems that cross-platform deployment of applications has finally been achieved. Due to the recency of WebAssembly as a compiler target, however, many toolchains (e.g., compiler backends for various languages, and competing WebAssembly runtimes) are not yet as mature as for existing targets. In fact, a recent work has shown that the semantics of programs developed for native binaries may not remain the same when compiled to WebAssembly [30]. We expand upon this work with the aim of identifying the root causes of these differences. This is particularly important from a security perspective: if there is a mismatch between the security guarantees of a native application and a WebAssembly application, it is crucial to be aware of them when porting applications to WebAssembly.

In particular, we investigate whether we can observe any difference in the outcome of the execution of C programs compiled to 64-bit x86 native code and to WebAssembly code. To this end, we compile 17 802 C programs containing common weaknesses to native binaries and to WebAssembly binaries. We run these programs and investigate whether any difference can be observed across their executions, meaning that the behaviour of the native executable differs from the corresponding WebAssembly one. We observe such differences in 4 911 programs, which may complicate porting C applications to WebAssembly. We investigate all cases that expose differences in behaviour and established their root causes. In total, we find three classes of root causes: a different standard library implementation, the lack of security measures in WebAssembly, and deviating semantics of the execution environments. We describe each root cause and illustrate examples that exhibit different behaviour. Importantly, we highlight the ones that may impact the security of WebAssembly applications. In summary, this work makes the following contributions:

- We identify three root causes for difference in execution of C programs compiled to WebAssembly and to native code.
- We discuss a set of examples exposing each root cause for a difference in behaviour, with a focus on the differences that are the most important from a security perspective.
- We publicly share the dataset of 4 911 C programs that exhibit divergent behaviour, together with the corresponding x86 and WebAssembly executables.[1]

The main implication of this work is to alert practitioners that porting a C application to WebAssembly may result in different program behaviour which can have an impact on the security of the application. Moreover, these findings encourage researchers to provide solutions to overcome these differences, as well as to investigate whether other differences in behaviour may still exist.

The remainder of this paper is organised as follows. In Section 2, we briefly provide the necessary background information on the WebAssembly language and its security model. Section 3 describes the selection of the dataset. Section 4 describes our approach to identifying and investigating behavioural differences between the WebAssembly and native binaries of a program. The results of our empirical analysis are detailed in Section 5 where we illustrate through a number of examples the differences that can arise when porting C applications to WebAssembly, and we explain their root causes. In Section 6, we explain the threats to validity of this work. We discuss related work in Section 7 before concluding the paper in Section 8.

---

[1]https://figshare.com/articles/dataset/SAC_2022_Dataset/17297477

## 2 BACKGROUND

We provide some background information on the WebAssembly language to streamline the understanding of what we discuss in this paper. We focus on the parts of the language that are relevant to our discussion. For a full reference on the language, we refer to its specification [26].

### 2.1 Execution Model

The execution model of WebAssembly is stack-based: instructions push values on and pop values from a value stack. Values can be stored in *local* variables and *global* variables, which are similar to registers. Arguments for function calls are passed through the stack; i.e., they need to be pushed on the stack beforehand. The return value of a function is the top value of the stack after executing the instructions in its body. Apart from the *value stack* which we described, there is no need to manually manage the *call stack* in WebAssembly for function calls. The actual jumping to and returning from function bodies is entirely managed by the runtime.

### 2.2 Memory Model

A WebAssembly application contains a single *linear memory*, i.e., a consecutive sequence of bytes that can be read from and written to by specific instructions. Using this linear memory properly is left to the program at hand. Hence, the runtime does not restrict in any way the usage of this linear memory. For example, there is no concept of page or segment in the linear memory, unless these are implemented in the program being executed directly.

### 2.3 Application Structure

A WebAssembly application is composed of, among other components, a number of functions. Functions have a number of parameters, and their body comprises a set of instructions. Some functions can be *exported*, and one such function can be specified as the entry point of the program. Exported functions are made available to the runtime. For example, programs compiled with the WebAssembly System Interface (WASI) expose a `_start` function, corresponding to the `main` function of a C program. This function will be called to start the program.

Functions have access to *local* variables, akin to local registers, in which they can store intermediary results. Function arguments can also be retrieved through these local variables.

An example function definition is the following:

```
1 (func $main (type 4)
2    (param i32 i32)
3    (result i32)
4    (local i32)
5    local.get 0
```

This function takes two 32-bit integers as parameters (Line 2), and returns a 32-bit integer (Line 3). When the function is called, the stack is initially empty and the parameters are stored in local variables 0 and 1. A third local variable is accessible as the function declares the need to access an extra local variable on Line 4. The function body consists of a single instruction on Line 5. This instruction accesses the value of the first local variable (i.e., the first argument of the function) and pushes it on the stack. After the last instruction, the function execution ends and the value remaining on the stack is the return value.

### 2.4 Example Instructions

We have seen the `local.get` instruction to access a local variable and push it on the stack. There exist many other instructions; those of interest are the following:

- `i32.const N` pushes constant N on the top of the stack.
- `i32.add` and `i32.sub` respectively add and subtract the two top values of the stack.
- `local.set N` pops the top value of the stack and sets the $N^{th}$ local variable to this value. `local.tee N` works similarly, but leaves the stack untouched.
- `i32.store` takes two values from the stack and stores the first value at the address pointed to by the second value in the linear memory.

Finally, note that there are variations of these instructions such as `i64.const`, which pushes a 64-bit integer value on the stack, or `i32.store8 offset=N` which stores a byte in the memory with the given offset.

### 2.5 Stack Memory Management

WebAssembly applications also have access to a number of *global* variables, which are accessed and modified similarly to local variables, but through the `global.get` and `global.set` instructions. C compilers like Clang rely on the first global variable, which we call `g0`, to model the stack memory of the C program they compile using the linear memory of WebAssembly: `g0` acts as the stack pointer. The portion of the linear memory starting at `g0` is therefore used to store stack-allocated data. A common pattern encountered in WebAssembly applications compiled from C that need to allocate space for stack data is therefore the following:

```
1 global.get 0
2 i32.const 64
3 i32.sub
4 global.set 0 ;; g0 becomes g0-64
5 [...] ;; function body
6 global.get 0
7 i32.const 64
8 i32.add
9 global.set 0 ;; restore value of g0
```

This excerpt takes the current stack pointer (`global.get 0`), decreases it by 64, and updates the stack pointer (`global.set 0`). This effectively allocates 64 bytes of data onto the stack. The rest of the function body can therefore store data in that portion of the memory. When the allocated memory is not needed anymore, the stack pointer is restored to its original value (Lines 6-9).

### 2.6 Security Model

The WebAssembly standard has been designed with security in mind, as evidenced among others by the strict separation of application memory from the execution environment's memory. Being executed in a sandbox, a compromised WebAssembly binary cannot compromise the browser or any other kind of runtime that executes the binary [2, 12].

Moreover, WebAssembly includes several features that aim at limiting the impact of a vulnerability being exploited. Unlike in x86, the return address of a function in WebAssembly is implicit and can only be accessed by the execution environment. This precludes among others return-oriented programming attacks, and reduces the potential for stack-smashing attacks. As another example, WebAssembly supports function pointers but only in so-called *indirect calls* for which the target function is determined by a statically-defined table. This again limits the range of control flow exploits that are possible through function pointers: only those functions that are declared to be possible targets of indirect calls can be called indirectly, which prevents calling arbitrary functions available in the application.

### 2.7 WebAssembly Vulnerabilities

Despite the sandbox in which they are executed and the overall security-minded design of the instructions of the language, WebAssembly binaries may still suffer from a number of security vulnerabilities [17] rendering them easily exploitable – sometimes easier than native code. For example, the use of critical functions such as eval or exec in a binary enables arbitrary code execution. The eval function enables evaluating arbitrary pieces of JavaScript on the browser executing the WebAssembly application. If one such function can be the legitimate target of an indirect call, and if the WebAssembly application is prone to buffer overflows that can rewrite function pointers for example, then the WebAssembly application may see its control-flow redirected to call it with untrusted data [18]. Although sandboxing precludes the need for provisions to protect the host from stack smashing and the likes, a WebAssembly application may still be prone to vulnerabilities that could allow arbitrary code execution of JavaScript for example [18].

## 3 DATASET

For our empirical analysis, we rely on the Juliet Test Suite 1.3 for C[2] of the *Software Assurance Reference Dataset* [3], released in October 2017 and which has been used to compare static analysis tools that detect security issues in C and Java applications [6]. It contains 54 484 test cases illustrating 118 different common weaknesses (CWE) that can arise in C programs. Each test case consists of a C program that contains a vulnerability, but no attempt to exploit it: for example, there is no attempt to reroute the control flow of the program by manipulating function pointers. As we focus on security-related differences, we chose this dataset because it exhibits common weaknesses that can cause security issues. We focus our analysis on programs that have deterministic behaviour, with the assumption that one expects such behaviour to match between the x86 binary and the WebAssembly application. Figure 1 depicts our dataset selection pipeline. We conduct our empirical analysis on a machine with an AMD Ryzen Threadripper 3990X 64-Core CPU (2.9 GHz) with HyperThreading and 256 GiB of RAM.

### 3.1 Categorisation of Test Cases

The test cases from this dataset are decomposed into 118 weaknesses, where for each weakness test cases are decomposed in further *categories*. All test cases within one such category rely on the same mechanism to trigger the weakness, but test cases within one category differ slightly in their control flow. For example, the test case called CWE121_Stack_Based_Buffer_Overflow_ _dest_char_alloca_cpy_01.c contains a stack-based buffer overflow (CWE 121), triggered by using a char buffer allocated with alloca and copied into with strcpy. This test case has 50 variations, each identified by a different number suffix and demonstrating different usages of the same functions to trigger the stack-based buffer overflow. Variations can, for instance, feature a more complex control flow stemming from additional conditionals. In total, there are 1258 categories in the dataset. We group programs per categories when we need to manually inspect them.

### 3.2 Pre-processing

In order to be able to compare runs of the program across the two configurations (WebAssembly and native), we remove any source of controllable non-determinism in the programs so that multiple runs are expected to produce the same results. To do so, we replace all calls to rand() with the constant 1. This choice is made because, after manual investigation of the programs in the Juliet Test Suite that call rand(), most are using the return value as a condition to execute the unsafe behaviour it contains. This does not mean however that all sources of non-determinism have been eliminated: programs can depend on uninitialised memory for example and exhibit non-deterministic behaviour. We will filter out such programs and do not consider them in our empirical analysis, as comparing non-deterministic programs based on their output will produce false positives when detecting differences of behaviour between the compilation targets.

### 3.3 Compilation of Test Cases

Each of the test cases can be configured to exhibit only the *safe* behaviour, only the *unsafe* behaviour, or *both*. We configure the dataset to include both behaviours when compiled. We are interested in particular in the unsafe behaviour, as we expect it to uncover more differences than the safe behaviours. However, in case there are no differences in the unsafe behaviour, there could still be potential differences in the safe behaviour too, which is why we consider both configurations at the same time. A program compiled with the configuration *both* will first execute its unsafe behaviour, and then execute its safe behaviour.

We compile each test case to WebAssembly and to 64-bit x86 with Clang v12.0.1[3], the latest version available when we performed our analysis in September 2021, with the default flags and the -O2 level of optimisations to reflect a real-world scenario.

We filter out programs that cannot be compiled to WebAssembly, as some programs depend on system-specific functions or on features that are not yet supported by WebAssembly, such as threads or sockets. In total, out of the 54 484 test cases, 19 534 can be compiled (36%).

---

[2]https://samate.nist.gov/SARD/testsuite.php

[3]All Clang flags and their default values are listed on the relevant documentation page: https://clang.llvm.org/docs/ClangCommandLineReference.html

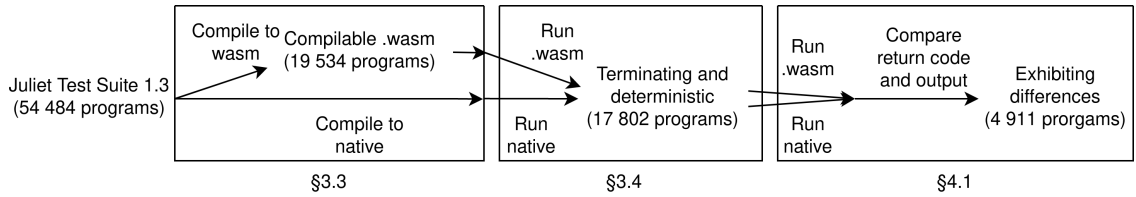

Figure 1: Dataset selection pipeline

## 3.4 Selection of Deterministic and Terminating Programs

Because we want to compare the results of running each binaries, we need to ensure that each binary produces deterministic results. As mentioned before, we did ensure that controllable sources of non-determinism (i.e., `rand`) are deterministic. The goal here is to remove programs that contain non-controllable sources of non-determinism. To that end, we first run each version of each program 10 times: if we notice that either the x86 executable or the WebAssembly executable is non-deterministic, we filter out this program. We manually inspected that each of the filtered out programs indeed had non-deterministic behaviour. Note that none of the programs in our dataset feature concurrency, as threads are not (yet) supported in WebAssembly.

Some of the programs in the Juliet Test Suite also expose non-terminating behaviour through programs that indefinitely print new output. Because the output of such programs will differ between their x86 and WebAssembly versions due to speed differences, we also exclude such programs from our empirical analysis. We identify them by running each program with a timeout of 100 seconds and by excluding those that fail to terminate within the time limit. This leaves us with 17 802 programs in total, which form our dataset.

## 4 RESEARCH METHOD

We describe the process that we followed in order to analyse the programs in our dataset.

## 4.1 Identifying Different Execution Behaviour

For each program, we have two executables: one compiled to 64-bit x86 native code, and one compiled to WebAssembly. We run each executable with a timeout of 100 seconds, and record its outcome in terms of return code (e.g., successful exit or crash) and in terms of standard output. The programs are executed on a machine running Linux 5.4.0, and the WebAssembly applications were run with the Wasmer 2.0.0[4] WebAssembly runtime, the latest version available at the time of writing. In total, our analysis spent 2h10 compiling to WebAssembly, 1h05 compiling to x86, 8h59 running WebAssembly applications, and 32 minutes running x86 binaries.

We then compare, for each program, the outcome of the native and WebAssembly binaries: in case there is a mismatch, this is an example of a difference in behaviour. We do notice a non-negligible subset of test cases for which both configurations produce different results: 28% of the dataset (4 911 programs) exhibit different behaviour exhibit different behaviour in their native version and in their WebAssembly version. We make this dataset of C sources, WebAssembly binaries, and x86 binaries publicly available.[5]

## 4.2 Manual Analysis

In total, there are 318 categories across 40 CWEs present in this set of programs that exhibit differences. We manually inspect 3 programs for each category, and check that all other programs in the same category which are marked as producing different behaviour, follow the same pattern of difference. For example, when 3 programs from the same category result in the same error in their native version but run to completion in WebAssembly, we check that, among the other programs that exhibit differences from that category, all their native version result in the same error, and all their WebAssembly version run to completion. Among a category, the core of the program remains the same but only differs in terms of control flow.

## 5 FINDINGS

We present the differences that we encountered, through various examples. As a summary of our findings, we extract the following three main root causes that explain these differences:

- *Differences due to a mismatch in the standard library implementations used for each binary.* When compiling to a native binary, the OS's implementation of the standard C library is used (in our case, `glibc`), while when compiling to WebAssembly it is the standard library provided by the WebAssembly System Interface (WASI), which is based on the `musl` standard library. Multiple differences occur from this use of a different implementation.
- *Differences due to missing security protections.* When something supposedly bad happens in native code, it can happen that the program crashes. This may be due to compiler protections or to OS-level protections. This is on purpose to avoid possible exploitation of this program, and can be detected by instrumentation added to the binary by the compiler, or directly at the hardware level. However, we notice that some of these protection mechanisms do not have an equivalent in WebAssembly, and the program therefore runs to completion.
- *Differences due to specifics of the execution environments of each binary.* There is behaviour that is specific to an execution environment and architecture, and the specifics of the WebAssembly platform can produce behaviour that is different from a native program executed on an 64-bit machine.

---

[4] https://wasmer.io/

[5] https://figshare.com/articles/dataset/SAC_2022_Dataset/17297477

We summarise how these root causes are encountered in Table 1.

## 5.1 Security-Critical Differences

We first focus on the differences that can have critical implications in terms of the security of an application, when compiled to WebAssembly.

*malloc/free implementation.* In order to compile C code to WebAssembly, we use the WebAssembly System Interface (WASI), which itself relies on the musl standard library instead of the standard library provided by the OS, which in our case is glibc. As a result, the implementation of standard library functions that are related to memory allocation and de-allocation see some difference in their behaviour, which may result in security concerns.

Consider the following code. A buffer of 100 bytes is allocated on Line 1 and initialized on Line 2. On Line 3, the pointer pointing to that buffer is incremented by 10, resulting in the pointer pointing to somewhere in the middle of the buffer afterwards. Finally, free is called with that pointer, which is an unsafe operation because free should always be called on the beginning of a dynamically allocated buffer.

```
1 char *data = malloc(100 * sizeof(char));
2 strcpy(data, SOURCE);
3 data += 10;
4 free(data);
```

When compiled to native code, this situation is properly handled by free: it prints the error free(): invalid pointer and aborts the program (SIGABRT). However, the WebAssembly application continues its execution after performing the free operation. As a result, due to a programmer error on Line 3, it could be the case that sensitive data remains accessible during the rest of the execution of the program.

This is an important difference that may extend beyond the use of musl, as many custom implementations of malloc and free for WebAssembly exist and are used in practice [15], and such differences in behaviour could be encountered in other implementations. In total, we encountered this difference in 259 programs. An example program that exhibits this difference is CWE761_Free_Pointer _Not_at_Start_of_Buffer__char_fixed_string_01.c.

*Missing Stack-Smashing Protections.* Code compiled to WebAssembly does not contain stack-smashing protections such as stack canaries. As a result, programs in which a stack smashing occurs and therefore may crash when executed natively, will see the corresponding overflow always undetected in WebAssembly. Consider the following code.

```
1 char * data;
2 char dataBadBuffer[50];
3 data = dataBadBuffer;
4 data[0] = '\0';
5 char source[100];
6 memset(source, 'C', 100-1);
7 source[100-1] = '\0';
8 for (i = 0; i < 100; i++) {
9     data[i] = source[i];
10 }
11 data[100-1] = '\0';
12 printLine(data);
```

This code allocates a destination buffer of 50 elements in the stack memory on Line 2, and a source buffer of 100 elements on Line 5, before copying the entire contents of the source buffer to the destination buffer through the loop on Line 8. However, because the destination buffer is too small, elements will be copied outside of the destination boundary.

We perform a manual inspection of the native x86 executable with radare2[6]. It operates as one would expect from the C source code, but it contains the following addition at the end of the compiled function:

```
mov rax, qword fs:[0x28]
mov rcx, qword [var_8h]
cmp rax, rcx
jne 0x12b7 ;; jump to final call instruction
add rsp, 0xc0
pop rbp
ret
call sym.imp.__stack_chk_fail ;; address 0x12b7
```

This is the code that checks the stack canary generated by the compiler and stops the program's execution when a stack overflow is detected. The canary is the value of the stack pointer before the execution of the function, and it is compared to the stack pointer after the execution of the function. If an inconsistency is detected, the execution of the program is aborted by calling __stack_chk_fail.

The C excerpt compiles to the following WebAssembly:

```
1 (func $main (type 4)
2   (param i32 i32) (result i32)
3   (local i32)
4   global.get 0
5   i32.const 64
6   i32.sub
7   local.tee 2 ;; l2 = g0-64
8   global.set 0 ;; g0 = g0-64
9   i32.const 0
10  ...
11  local.get 2 ;; [g0]
12  i32.const 67 ;; ['C', g0]
13  i32.const 99 ;; [99, 'C', g0']
14  call $memset
15  local.tee 2
16  i32.const 0
17  i32.store8 offset=99
18  ...)
```

We notice that the source buffer has been inlined; the generated code operates directly on the destination buffer. Line 8 moves the stack pointer (g0) up by 64 bytes to allocate stack space for the destination buffer. The memset call on Line 14 fills it with 99 C characters. Afterwards, Line 17 writes the final null character to position 99 of the destination buffer. Note that because the destination buffer has only been allocated 64 bytes, these operations will result in a stack overflow. However, we do not observe the presence of a stack protection mechanism in the WebAssembly code, and executing

---
[6]https://rada.re/n/

Table 1: Breakdown of root causes and actual differences.

| Root cause | Due to | Programs affected |
|---|---|---|
| Different standard library | | 3574 |
| | Wide characters | 3253 |
| | malloc/free | 259 |
| | puts | 36 |
| | printf | 26 |
| Security protections | | 769 |
| | Stack smashing | 626 |
| | Memory protections | 143 |
| Execution environment | | 444 |
| | Uninitialised data | 382 |
| | Size of pointers | 26 |
| | Size of numbers | 18 |
| | OS' environment | 18 |
| | Memory layout | 18 |

the WebAssembly application will not result in a crash, letting the overflow occur silently.

We observe this difference in behaviour in 626 programs. An example program that exhibits this difference is `CWE665_Improper _Initialization__char_cat_01.c`.

*Missing Memory Protections.* A number of differences are caused by memory protections being present in x86 but not in WebAssembly. As a result, a native executable will crash with a segmentation fault (SIGSEGV), usually being detected at the hardware level. However, in WebAssembly, there is no equivalent protection as there is no concept of page or segment in the linear memory. The linear memory is a contiguous block of bytes, and there is no restriction in reading from or writing to it.

For example, buffer underwrites are a situation where data is copied *before* its destination buffer. In native code, underwrites can be detected at the hardware level, or through some form of bounds checking, and a program performing an underwrite will often crash with an address boundary error. However, in WebAssembly, such underwrites remain undetected and the program continues to run.

This situation is encountered in the following example. A buffer of 100 bytes is allocated and filled in Line 1-4. On Line 4, the `data` variable points to 8 bytes *before* the buffer, and therefore to an invalid location. Line 9 copies a source buffer of 100 bytes at the location pointed by `data`, resulting in a buffer underflow.

```
1 char *dataBuffer =
2   (char *)alloca(100*sizeof(char));
3 memset(dataBuffer, 'A', 100-1);
4 dataBuffer[100-1] = '\0';
5 data = dataBuffer - 8;
6 char source[100];
7 memset(source, 'C', 100-1);
8 source[100-1] = '\0';
9 strcpy(data, source);
```

We observe this difference in behaviour in 143 programs. An example program that exhibits this difference is `CWE124_Buffer _Underwrite__wchar_t_alloca_cpy_01.c` and

### 5.2 Non-Security-Critical Differences

We briefly present other behavioural differences we encountered. These differences are of less importance when it comes to the security of the application, but are nonetheless useful to be aware of when porting a C application to WebAssembly.

*Different Standard Library Implementation.* As mentioned previously, programs compiled to WebAssembly with the WebAssembly System Interface (WASI) rely on the `musl` libc implementation. In contrast, when they are compiled to native code in our setup, they rely on `glibc`. This difference in standard library implementation results in several behavioural differences. Besides the different behaviour of `malloc` and `free`, which we consider a critical difference and covered in the previous section, a number of other functions exhibit differences. In our dataset, we found behavioural differences that can be linked to the following elements of a standard C library.

- *Wide character's mode for* `wprintf` (3253 programs). In native code, `wprintf` has the default behaviour of not printing anything to the console unless `fwide(stdout, 1)` has been called before. However, in WebAssembly, we observe that by default, `wprintf` does print to the console, resulting in the output to differ compared to the native executables. An example programt htat exhibits this difference is `CWE126_Buffer_Overread__CWE170_wchar_t_loop_01.c`
- `puts` *return value* (36 programs). Functions `puts` and `fputs` return a non-negative number upon success. The return value however depends on the libc implementation: in `musl`, 0 is returned upon success, while in `glibc`, a positive number is returned. This results in a divergent behaviour for some CWE253 benchmarks (*Incorrect Check of Function Return Value*), which check whether the return value of `fputs` is 0. An example program that exhibits this difference is `CWE253_Incorrect_Check_of_Function_Return_Value_ _char_fputs_01.c`.
- *Missing arguments to* `printf` (26 programs). Calls to `printf` with missing arguments behave differently. For example, `printf("%s")` prints some "garbage" when executed on x86, but prints `(null)` when executed in WebAssembly, indicating a different implementation of `printf` from musl. An example program that exhibits this difference is `CWE134 _Uncontrolled_Format_String__char_console _vfprintf_44.c`.

*Security Protections.* We covered the critical differences in the previous section. We did not encounter any difference in terms of security protection that we considered non-critical.

*Semantic Differences in Execution Platforms.* The remaining differences can be traced back to the WebAssembly and native execution platforms.

- *Size of pointers* (26 programs). On a 64-bit machine, pointers are 8 bytes long, while in WebAssembly they are 4 bytes. Hence, the return value of `sizeof(void *)` differs between the two, resulting in observable differences in execution. An example program that exhibits this difference is `CWE789 _Uncontrolled_Mem_Alloc__malloc_char_fgets_01.c`.
- *Different number sizes* (18 programs). This is related to the previous point: `long` does not have the same size in WebAssembly as on a 64-bit machine. Therefore, we observe differences for example in the return value of `strtol` when it results in an overflow and therefore returns `LONG_MAX` as the default value, which is set to $2^{63} - 1$ on a 64-bit machine, while in WebAssembly it is set to $2^{32}-1$. An example program that exhibits this difference is `CWE391_Unchecked_Error _Condition__strtol_01.c`.
- *Uninitialised data behaviour* (382 programs). Relying on uninitialised data in C is considered as an undefined behaviour. In native code, accessing uninitialised data may trigger a `SIGSEGV` error or result in garbage being treated as data. In WebAssembly however, the linear memory is initially filled with 0s. As an example, printing a string that is not null-terminated in native code may print garbage after the string, while in WebAssembly, the byte that follows the string often is 0, which acts as the null terminator. In other cases, when the WebAssembly application reads from outside of the expected bounds, it can reach data that has already been written to, while a native application will crash or reach different data. We therefore encounter differences in output. Similarly, accessing a pointer whose value is `NULL` can also trigger a `SIGSEGV` error or use uninitialised data. In WebAssembly however, in both situations the program continues its execution and uses data that is set to 0. The following example demonstrates this: `*pointer` is 0 as it is not initialised, hence printing `*data` prints a string that contains only the character \0.

```c
double **pointer =
    (double **)alloca(sizeof(double *));
double *data = *pointer;
printDoubleLine(*data);
```
An example program that exhibits this difference is `CWE457 _Use_of_Uninitialized_Variable__char_pointer_01.c`.
- *Different execution environments* (18 programs). We observed some differences due to the execution environment being different in WebAssembly and in native code. For example, the `getenv` function can be used to access the executable's environment variables. In the native binary, the environment is inherited from the process that launched it, and it contains environment variables defined by the user (e.g., `PATH`). With wasmer however, the environment is initially empty unless specified otherwise. An example program that exhibits this difference is `CWE526_Info_Exposure_Environment _Variables__basic_01.c`.

### 5.3 Discussion

We observed multiple differences in behaviour between a C application compiled to native code and compiled to WebAssembly. We see that porting a C application to WebAssembly may not be as easy as switching compiler flags. Although it is not always clear who is responsible for these differences (i.e., are they the responsibility of the programmer, the standard library, the compiler, or the runtime?), it is important to identify them. Moreover, we believe that addressing them earlier in the development pipeline is to be preferred: having the same security measures put in place by the compiler across compilation targets is a desirable property.

*Differences with security implications.* Most importantly, there are a number of differences that may have a serious impact on the security of the program. Even though WebAssembly has been designed with security in mind so that for example, control-flow hijacking attacks through e.g., return oriented programming are impossible, and even though WebAssembly is executed in a sandboxed environment, there are still some concerns. For example, the design documents of WebAssembly state that "common mitigations such as [...] stack smashing protection (SSP) are not needed by WebAssembly programs", while clearly WebAssembly applications can be the target of overflow-based attacks as we have seen with the *missing stack-smashing protection* differences: stack smashing may cause a native program to crash due to compiler protections, while the program will continue its execution in WebAssembly. In practice, exploiting such a vulnerability may be more complicated in WebAssembly than in a native binary: the call stack does not reside in WebAssembly's linear memory where the overflow happens, and typical stack smashing exploits are therefore prevented. However, the buffer overflow goes undetected and may overwrite data used later in the program, potentially controlling the future of the program's control flow. Even though the program remains sandboxed in its environment, the control or data flow of the program may be exploited by an attacker. This is particularly important for server-side WebAssembly applications, for example executed on the wasmer environment, or integrated in Node.js applications. For example, Node.js applications relying on vulnerable WebAssembly code could then be prone to remote code execution attacks [17]. Similarly, WebAssembly applications do not feature the same memory protection as native binaries, resulting in different outcomes in case of, for example, buffer underwrites. A final critical difference is related to the memory allocation functions. It has been observed that a non-negligible set of WebAssembly applications relies on non-standard allocators: no less than 11 allocators have been found in the wild by Hilbig et al. [15], with 38% of the programs relying on a custom allocator. The use of a non-standard allocator can be motivated by the importance of minimizing code-size when deploying applications on the Web, and by the use of toolchain-specific allocators, e.g., provided by Emscripten, the Go, Rust, or AssemblyScript toolchains. Such differences may be observed with other memory allocators that are used in practice when compiling to WebAssembly.

*Other differences without security implicitaions.* There are a number of additional differences to be aware of when porting an application to WebAssembly, even though these have a lesser impact on the security of the application. These are related to the use of different standard libraries and to the different semantics of the execution platforms. Regarding the standard library, musl is an established alternative to glibc, which has seen its first stable release 10 years ago and has been used in many Linux distributions as the default standard library. However, we still observe non-negligible differences of behaviour with glibc. This indicates that one has to remain careful when porting an application to WebAssembly.

*Reliance on undefined behaviour.* An important aspect is that the code discussed here sometimes relies on undefined behaviour. Even though this means that compilers are allowed to transform such programs and that there is no guarantees about their execution, it is important to know how such behaviour may differ across different compilation targets.

As a result from our empirical evaluation, we see that besides a number of non-critical differences in the execution of C application compiled to WebAssembly with respect to their native code execution, there are three differences that have an impact of the security of a C application when ported to WebAssembly.

## 6 THREATS TO VALIDITY

We identify three main threats to validity. First, a threat to external validity is that we rely on an existing test suite of programs with known vulnerabilities. The set of vulnerabilities included in this test suite may not be exhaustive, and the root causes we identified may therefore not be exhaustive either. The latest release of the Juliet Test Suite dates from 2017, and there could be other vulnerabilities or patterns to express them that are not present within the test suite. For instance, one aspect missing in the programs we inspected is the use of function pointers. We selected all programs that could be compiled to WebAssembly and that exhibited deterministic behaviour when ran multiple times, in order to enable the comparison. There could however be differences between the behaviour of the WebAssembly binary and the native binary for non-deterministic programs, although these require a much more careful analysis and are beyond the scope of this paper. Similarly, we replaced calls to rand by the constant 1 to avoid unnecessary non-determinism. This has no impact on the differences that we have encountered (i.e., no false positives) nor on the identification of their root causes, but this does imply that some differences may have been missed (i.e., potential false negatives).

A threat to internal validity is the fact that we focus on inconsistencies which are exhibited through differences in the program outcome, both on the level of the return code and of the standard output. Programs may produce the same output, yet have different semantics depending on the execution platform. Identifying different behaviour for programs that result in the same outcome is something that requires a thorough analysis of programs which we leave for future work.

Finally, another threat to internal validity regards the setup of our empirical analysis. We performed our analysis using Clang to compile programs in WebAssembly, while it is also possible to compile them using GCC with the Emscripten toolchain[7]. We leave a comparison to the results of GCC for future work. Moreover, all executables were run on a Linux platform with the wasmer runtime; results may differ on another architecture, platform or with another runtime. We performed our analysis using -O2 as the optimisation level. Repeating the analysis under alternative compiler configurations could highlight more root causes of divergence.

## 7 RELATED WORK

The number of security issues has increased in the past years, and only a tiny group of developers are responsible to address such issues [4]. Therefore, researchers and practitioners have put great effort in identifying and resolving security risks [8–11, 14]. In the following, we only discuss the most relevant literature to this work.

### 7.1 WebAssembly Applications Security

WebAssembly has been designed from the start with a focus on security [2]. Despite this focus, WebAssembly binaries may still suffer from a number of weaknesses. Lehmann et al. [17] demonstrated through a number of exploits that WebAssembly binaries can be more easily exploitable than native binaries. For example, the presence of critical functions exported from the environment such as eval or exec may enable arbitrary code execution in the browser. These security risks have also been identified by McFadden et al. [18].

A recent study of the usage of WebAssembly in a wide range of sources (websites, code repositories, and package managers) [15] demonstrates that around two thirds of WebAssembly binaries are developed in memory-unsafe languages, and are therefore prone to vulnerabilities that affect the source language. Romano et al. [25] performed an empirical study of bugs in WebAssembly compilers, based on existing bug reports. Our method is not focused on already fixed compiler bugs, but instead enables the *detection* of inconsistencies that could be potential bugs.

### 7.2 Hardening WebAssembly

There have been numerous works related to harden WebAssembly applications in order to improve their security, by extending the language, improving compilers and tools, or by improving runtimes.

Disselkoen et al. propose an extension to WebAssembly that allow developers to encode C and C++ memory semantics in WebAssembly, by reifying segments and handles in the language [7].

---

[7]https://emscripten.org/

Swivel [22] is a new compiler framework that hardens WebAssembly binaries against Spectre attacks, which can compromise the isolation guarantee of WebAssembly. Arteaga et al. propose an approach to achieve code diversification for WebAssembly [5], by generating multiple programs variants from an input program. This mitigates attack vectors on vulnerabilities that a application can have, after having been compiled to WebAssembly. Namjoshi et al. present a self-certifying compiler for WebAssembly [21], so that the optimisations performed during compilation are generated with proofs of their correctness.

Static analysis can also help identify potential security risks in an application. Wassail [28] is a static analysis framework for WebAssembly that has been used to build an information flow analysis [27] in order to detect higher-level security concerns such as leaks of sensitive information.

At the level of the runtime, Ménétrey et al. [19] present a *trusted* runtime for WebAssembly by executing the runtime inside a trusted execution environment such as Intel SGX, thereby diminishing the potential impact of vulnerabilities. Nieke et al. [23] also rely on SGX to improve host security in the context of using WebAssembly binaries for edge computing. Solutions based on mechanisms similar to SGX may however suffer from some limitations when it comes to input/output or when it is necessary to exit from the boundaries of the enclave.

### 7.3 Identifying Language Bugs and Inconsistencies

Our method for identifying the difference between multiple compiler configuration is inspired by compiler fuzzing. Csmith [31] performs fuzzing of compilers by generating C programs, compiling them with several compilers, and observing their result: any difference in output is likely a bug. Similarly, Midtgaard et al. [20] discover bugs in the backends of the OCaml compiler, by generating source code which is then compiled with two different backends. If a difference in output is observed, a potential bug is found. In our case, we also target two backends of the same compiler, but we rely on a predefined set of input programs by using the Juliet Test Suite for C instead of generating these programs. Stiévenart and Madsen [29] apply the same idea, but only to one configuration at a time: given a program that must adhere to a given behaviour, if, during its execution, the program does not yield the expected behaviour, a language bug is identified. The idea of generating programs has been applied to WebAssembly by Perényi and Midtgaard [24]: generating WebAssembly applications and running them with multiple runtimes enables the detection of disagreements between these runtimes.

When only one compiler configuration is available, one can rely on Equivalence Modulo Input (EMI) instead [16]: two programs that differ in their source code but not in the expected behaviour are compiled and run, and if any difference is observed, a compiler bug is potentially found.

## 8 CONCLUSION

We investigated whether differences can arise between the execution of a C program compiled to WebAssembly and the execution of that program compiled to another target such as x86. We studied the consequence of such differences, and whether they may have a critical impact on the security of an application.

We have observed that, out of 17 802 C programs containing known weaknesses, 4 911 differ in outcome when their WebAssembly and their x86 binary is executed, either by printing different output, or by differing in their return code. We manually inspected these differences and observed that they are the result of three main root causes. First, the use of a different standard library often results in minor issues, but also in critical issues from a security perspective due to the different guarantees provided by the memory allocators. Second and more importantly, there are critical differences due to security protections, particularly stack-smashing and memory protections, that are missing in WebAssembly. Consequently, a native program that would potentially crash when encountering unsafe behaviour, may still run to completion in WebAssembly. Therefore, applications that were protected from certain vulnerabilities when compiled to native code can become insecure when compiled to WebAssembly. Finally, there are minor differences that are due to the semantics of each execution platform, such as the size of numeric data types or behaviour with respect to uninitialised data.

We believe that these observations are important for practitioners when porting C programs to WebAssembly. Future research is needed to scrutinise these differences and resolve them to enable the secure and painless deployment of WebAssembly binaries compiled from C programs. Finally, follow-up research may explore whether more of such differences can occur in other languages and environments.